\documentclass[11pt,a4paper]{article}

\renewcommand{\author}{Emil Khalisi}
\newcommand{\titel}{The Solar Eclipse of the Xia Dynasty: A Review}

\renewcommand{\date}{28th June 2020}

\usepackage{hyperref}
\hypersetup{pdfauthor={Emil Khalisi}}
\hypersetup{pdftitle={\titel}}
\hypersetup{pdfkeywords={solar eclipse, astronomical dating, chronology,
   Xia dynasty, ancient China, agglomeration of planets, Shujing, Zhong Kang}}

\usepackage[T1]{fontenc}        % ESSENTIELL: Richtig trennen, Umlaute benutzen
\usepackage{mathptmx}           % Grundschriftart Times + Mathematiksatz
\usepackage{pifont}             % Zapf-Dingbats + Symbol aktivieren
\usepackage[scaled=0.92]{helvet} % Grundschriftart Helvetica, serifenlos
\usepackage[british]{babel}     % ! falls Fehlermeldung: .aux-Datei loeschen
\usepackage{amsmath}            % Erweiterungen fuer mathematische Formeln
\usepackage{microtype}          % Perfektionieren von Randausgleich, Skalierung, Zeilenumbruch

\usepackage{multicol}           % Mehrspaltiger Text (Begleiter, Ch. 3.5)
\usepackage{multirow}           % Mehrspaltiger Text (Begleiter, Ch. 3.5)
\usepackage{eurosym}            % Laden des Euro-Symbols
\usepackage{units}              % Einheiten und Brueche im nice-, text- und math-mode

\usepackage{setspace}           % fuer spacing-Umgebung (*vor* KOMAoptions setzen!)
\usepackage{calc}
\usepackage[a4paper]{geometry}  % NICHT verwenden mit DIV-Option !!!
\usepackage{scrextend}
\changefontsizes{10pt}
\usepackage{titlesec}
\titleformat*{\section}{\large\bfseries}
\titleformat*{\subsection}{\normalsize\bfseries}

\usepackage{graphicx}           % Unmittelbare Umwandlung in pdf-File
\usepackage{float}              % Organisieren d. Float-Objekte (Figs, Tab., Captions)
\usepackage{caption}            % Non-float Objekte
\usepackage{fancyhdr}
\renewcommand{\headrulewidth}{0.4pt}
\usepackage{colortbl}           % fuer \rowcolor noetig!
\definecolor{grey20}{RGB}{208,208,208}
\usepackage{lineno}             % Nummerierung der Zeilen

\begin{document}

%%%%%%%%%%%%%%%%%%%%%%%%%%%%%%%%%%
%%%%%   Abstract for arXiv   %%%%%
%%%%%%%%%%%%%%%%%%%%%%%%%%%%%%%%%%

\fancyhead{}
\fancyhead[LO]{%
   \footnotesize \textsc{German Version:} \\
     see Habilitation at the University of Heidelberg
}
\fancyhead[RO]{
   \footnotesize {\tt arXiv: 2006.04674 [physics:hist-ph]}\\
   \footnotesize v2: {28 June 2020}%
}
\fancyfoot[C]{\thepage}

%%% Umschalten von 2-column auf 1-column ohne pagebreak:
%%% ----------------------------------------------
\twocolumn[
\begin{@twocolumnfalse}

\section*{\centerline{\LARGE \titel }}

\begin{center}
{\author \\}
\textit{69126 Heidelberg, Germany}\\
\textit{e-mail:} \texttt{ekhalisi@khalisi.com}\\
\vspace{\baselineskip}
\end{center}

\begin{abstract}
\changefontsizes{10pt}
\noindent
We present a review on the renowned solar eclipse in the Chinese
book \textit{Shujing} describing the oldest account of this kind.
After balancing the arguments on the time, place, and the celestial
stage, we offer a new scenario.
The path of totality on 15 September 1903 BCE traversed Anyi,
the assumed capital of the Xia dynasty.
The date matches remarkably well the chronological order of two
other incidents of mythological rank:
the closest-ever agglomeration of the naked-eye planets and the
Great Flood under Emperor Yu.
Still, this eclipse does not remove all questions about the
historical circumstances given in the account.\\

\noindent
\textbf{Keywords:}
Solar eclipse,
Astronomical dating,
Chronology,
Xia dynasty,
Ancient China

%\vspace{\baselineskip}
%\noindent
%{\small Received: ***. Accepted: ***}

\end{abstract}

\centerline{\rule{0.8\textwidth}{0.4pt}}
\vspace{2\baselineskip}

\end{@twocolumnfalse}
]
%%% ----------------------------------------------

%%%%%%%%%%%%%%%%%%%%%%%%%
%%%%%   Main text   %%%%%
%%%%%%%%%%%%%%%%%%%%%%%%%

\section{Introduction}

The solar eclipse in the reign of the Xia Emperor Zhong Kang is
one of the best known among the historical accounts on eclipses.
The event is associated with the dreadful fate of two astronomers,
named He and Ho, who were not prepared for it and therefore
punished.
This oldest account known, held in the classical work
\textit{Shujing} (also \textit{Shu Ching} or \textit{Shoo King}),
surpasses any other by more than 800 years.
A lot of information about this eclipse is rather legendary than
steady, as will be shown.

To understand the connection between the mythological character,
its fame, and the reality, one has to trace down the history of the
sources as well as the political importance of astronomy in the
days of ancient China.
This article tries to review the passed down facts.
First, we briefly examine the origins of the \textit{Shujing}
and the derivative descriptions and check for their conclusiveness.
We quote the relevant paragraphs from these works.
After summarising previous analyses, we compare the proposed dates,
and finally suggest a new solution that would be consistent with
two other incidents of that obscure era.
In conclusion, we wish to add the new date to the evidence in
favour of the historicity of delineated events, although the
account itself is of doubtful nature.

An essential quantity for studying historical eclipses is the
``clock error'', $\Delta T$.
It denotes the deceleration parameter of the earth's rotation:
the difference between a theoretically uniform time scale
(Ephemeris Time) and the present time (Universal Time).
This issue will not be covered here, but we touch on the problem
at the very end pointing out that our result perfectly fits the
average value of the extrapolated $\Delta T$.

Dates will be given historically, i.e.\ omitting the year zero.
For example, the year ``1903 BCE'' is ``-1902'' of the astronomers.

%%%%%%%%%%%%%%%%%%%%%%%%%%%%%%%%%%%%%%%%%%%%%%%%%%%%%%%%%%%%

\section{At the dawn of the Xia dynasty}

The Xia dynasty is the first kingdom that controlled a major domain
of tribes in China.
According to traditional texts there existed ``Five proto-Emperors''
with some kind of godlike status before the Xia.
Some knowledge about these legendary Emperors helps to understand
the events preceding the solar eclipse.

\subsection{On the ancient text sources}

Almost all we know about the solar eclipse at the dawn of Chinese
history goes back to the \textit{Shujing},
the ``Book of Historical Documents''.
Its origin is particularly complex.
Apparently written by various authors, it is attributed to the
philosopher Confucius (551--479 BCE).
The book contains speeches, royal decrees, proclamations, and
appointments of high officials since the foundation of the kingdom.
Embedded into an elegant language, many important events are
conveyed there.
Starting with those pre-dynastic emperors, the oldest documents are
estimated at 2400 BCE, the latest at the 7th century BCE
\cite{chi_fotheringham_1933}.
The collection contains no continuous historical narrative and no
attempt at chronology.

The original of the \textit{Shujing} was destroyed in the burning
of books in 213 BCE, together with many other works.
This event represents a pivotal moment in the history of China,
as the monarch of the short-lived Qin dynasty performed various
acts against Confucianism and other beliefs.
He wanted to reform the state and re-write history from scratch.
The destructive operation was to leave behind a great impact on the
society and culture.
In spite of the immense loss of precious historical treasures, a
few changes led to some positive aftermath in the following Han
dynasty.
However, the burning of the books is the main cause for our lack of
knowledge about the very early times.

Attempts were made to recover the book.
Fragments and a table of contents escaped the destruction.
An incomplete copy of the \textit{Shujing} could be found about
25 years later.
It is said to have consisted of 28 or 29 out of the original 100
chapters, but there was no mention of an eclipse \cite{chi_brown}.
Then, a ruler in the 4th century AD ordered to restore the book,
i.e.\ more than a half thousand years after the loss.
Some citations were gathered from other books, which one could get
hold of, other parts were possibly replenished.
No matter whether or not the restoration could succeed, it seems
obvious that gaps would be inevitable.
We might call it a forgery \cite{chi_fotheringham_1933}.
But the biggest defect concerns, from our present point of view,
that the unknown restorer did not specify when he was filling in or
when conjecturing a possible story, and even not which sources he
followed.
It seems that he just headed for the result rather than cared for
the route towards it.

Most scholars of the 19th century put forward that the documents
were selected by Confucius not for their historical importance
but for teaching morality
\cite{chi_schlegel-kuehnert, chi_russell_1895, chi_chambers}.
The lapse of the astronomers He and Ho did \emph{not} consist of
the missing prediction of an eclipse but the disregard of their
duty.
The royal officers were to superintend the customary rites.
These rites were to apply the royal regulations like commissioning
archers, beating of drums, changing clothes, igniting incenses,
and something like that.
The emperor should perform ceremonies and carry out certain acts
to ``prevent the world from destruction''.
The Chinese belief was that a dragon or monster would approach the
luminary and threatened to devour it.
The people were to dispel the dragon by making noise with all
kinds of equipment.
Fortunately, they succeeded each time.

The existence of royal instructions for solar eclipses presupposes
some knowledge about the phenomenon.
People could not forecast it, but they knew how to cope with it.
There were persons in charge and they had to administer the
procedure.
Though nobody could predict dates, people could react accordingly
as soon as the ``dragon'' was advancing.

Besides the \textit{Shujing} there are two more passages from
other sources referring to most likely the same eclipse.
The second one emanates from the \textit{Annals of the Bamboo
Books}.
Before the invention of paper in the first century BCE, bamboo was
a common writing medium, and strips of typical length of 50 cm were
tied together to form a book.
This chronicle escaped the burning of books just by chance, because
it was entombed with the relicts of a king who had died in 296 BCE.
It was discovered, together with other scripts, by tomb raiders in
281 AD \cite{pankenier_1980}.
Some parts of the bamboo strips were destroyed but the remaining
were restored and copied several times.
The \textit{Bamboo Annals} also report about those pre-dynastic
emperors as well as the dawn of the Chinese culture.
The history ends with the aforementioned king who kept it in his
tomb.

The third text is by the historian Zuo Qiuming (ca.\ 556--451 BCE).
He lived in the state of Lu which was the home of Confucius.
Some historians say that he was a student of the latter and did not
make as large an appearance as other students, thus, he is less
prominent.
Other historians just tell that he was contemporary to Confucius.
Zuo Qiuming is best known for his commentary on the ancient
chronicle \textit{Chunqiu} or ``Spring and Autumn Annals''
--- not to be confused with the \textit{Bamboo Annals} above.
Written in narrative style, it covers the period from 722 to
468 BCE and focuses mainly on political, diplomatic, and military
affairs from that era.

%%% ONLY FOR ARXIV-VERSION:
%\pagestyle{plain}
\renewcommand{\headrulewidth}{0pt}
\fancyhead{}
\fancyhead[CE, CO]{\footnotesize \itshape E.\ Khalisi (2020): \titel}

%%%%%%%%%%%%%%%%%%%%%%%%%%%%%%%%%%%%%%%%%%%%%%%%%%%%%%%%%%%%

%%%%% Figure: Great Flood
%
\noindent
\begin{figure}[t]
\includegraphics[width=\linewidth]{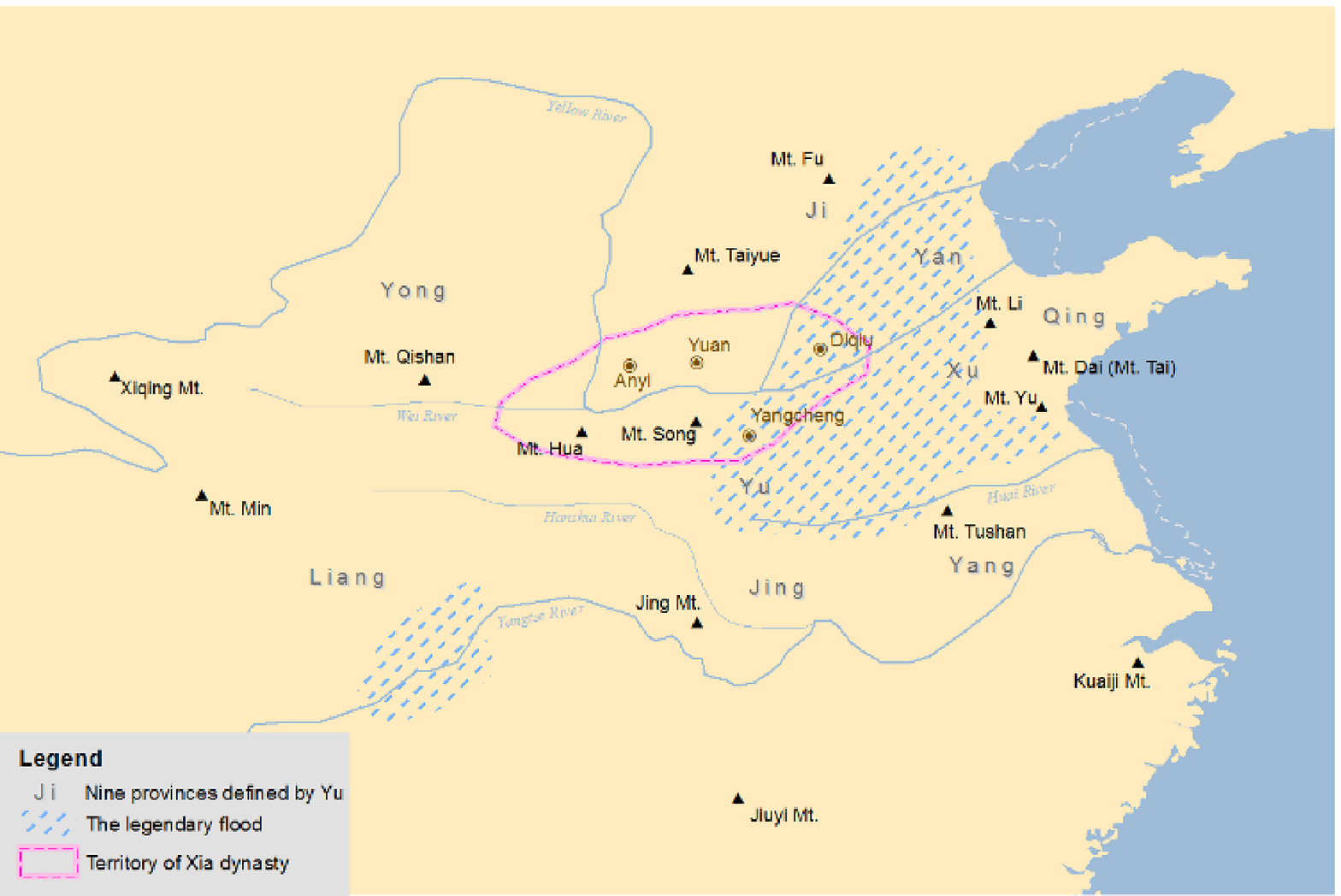}
\caption{Approximate territory of the Xia Dynasty and the inundated
    areas \cite{chi_w-en}.}
\label{fig:greatflood}
\vspace{-0.5\baselineskip}
\end{figure}

%%%%%%%%%%%%%%%%%%%%%%%%%%%%%%%%%%%%%%%%%%%%%%%%%%%%%%%%%%%%

%%%%% Table: Xia Dynasty
%
\begin{table*}[t]
\centering
\caption{Proposals for the reign of the Xia dynasty.}
\label{tab:xiadynastie}
\begin{tabular}{r@{ -- }llc}
\hline
\rowcolor{grey20}
\multicolumn{2}{l}{\cellcolor{grey20}{Time span [BCE]}} &
       Method of dating & Ref. \\
\hline
2205 & 1766 & Traditional chronology & \cite{chi_pang_1987} \\
2200 & 1675 & Timeline of rulers (Wikipedia) & \cite{chi_w-en} \\
2183 & 1751 & Solar activity / method unclear (1980) & \cite{chi_wang-siscoe} \\
2100 & 1800 & Radiocarbon method (1992) & \cite{chi_fairbank} \\
2070 & 1600 & Xia-Shang-Zhou-Project (1996--2000) & \cite{chi_liu_2002} \\
1989 & 1558 & Annals of the Bamboo Books ($\approx$ -300) & \cite{chi_legge} \\
%%2000 & 1520 & (no information on method, 1980) & Needham (1957)  \\
%%1953 & 1555 & (no information on method, 2013) & Pankenier (2013) \\
1953 & 1550 & Planetary grouping + average regency & \cite{chi_pang_1987} \\
%%1914 & 1600 & Kalenderzyklus (1990) & \cite{chi_nivison-pang, pang-yau_1996} \\
1900 & 1550 & Geologic stratigraphy (2016) & \cite{chi_wu-etal_2016} \\
\end{tabular}
\end{table*}

%%%%%%%%%%%%%%%%%%%%%%%%%%%%%%%%%%%%%%%%%%%%%%%%%%%%%%%%%%%%

\subsection{The Great Flood}

The introducing chapter of the \textit{Shujing} begins with a
glorious regent Yao who is hardly datable.
Some put him in the middle of the 24th century BCE, others to the
end of the 22nd century BCE.
For example, the translator of the \textit{Shujing}, James Legge
(1815--1897), fixed his accession to 2357 BCE,
while the astronomer Ludwig Ideler (1766--1846) did that to
2163 BCE \cite{chi_legge}.

Yao was one of those Five Great Emperors before the first
hereditary dynasties.
According to the \textit{Bamboo Annals} he died at the age of 119.
Before him there were other supremes with reigns of more than
10,000 years.
Such statements deprive of dating in general.
Any information from the early times remains extremely fuzzy till
$\approx$770 BCE when the verifiable history starts.
If old dates could be interpreted anyhow, then just by making use
of celestial happenings.

The first astronomical remark in the \textit{Shujing} is that
Yao ordered to determine the cardinal points of the sky and founded
the calendar ``at the beginning of time''.
He was advised by two astronomers named He and Ho.
They were to observe carefully each celestial part represented by
the solstices and the equinoxes, respectively
\cite{chi_russell_1895}.

During the reign of Yao there would have been a tremendous flood,
also of mythological rank (Figure \ref{fig:greatflood}).
The passage referring to it appears later on the documents, and a
similar account was given by other historians between the 4th and
1st century BCE \cite{chi_pang_1987}.
If one gives credence to the records, the deluge was the most
severe since the ice age.
Geologists assume a massive earthquake in Eastern Tibet that led
to landslides, dammed lakes as well as redirections of rivers.
It would have modified the landscape and altered the agricultural
conditions.
The disaster has been dated at 1922$\pm$28 (1$\sigma$) BCE utilising
the radiocarbon method on skeletons of three cave dwelling
human victims \cite{chi_wu-etal_2016}.
The 95\% confidence interval ranges from 1976 to 1882 BCE.

This geologic study provoked controversy among scholars.
We will not re-discuss the pros and cons here,
see \cite{wu-etal_2017-response} and references therein.
Another work argues that the same sediments, that have been used
in that analysis, indicate rather two separate landslides
having occurred at 8,300 and 6,300 BCE, respectively
\cite {zhang-etal_2019}.
The two dammed lakes gradually shallowed and then disappeared,
but both preceded the Great Flood of the Xia by 1400 years,
at least.
The geomorphic events would be unrelated to the historical events
in the writings.

The legends say that the Great Flood lasted for two generations,
putatively two decades, at least,
till Yao's second successor, the engineer Yu, attained a solution
to the problem.
After ``taming the waters'', he was conventionalised as a hero,
then ennobled as emperor and founder of the first dynasty,
the Xia.
He was given the epithet ``the Great''.
Furthermore, the calendar ought to have started with him (again?)
upon a celestial sign which will be outlined below.
Yu is said to have ascended 100 years after Yao.
This contradicts the statement on the inundation of ``two decades'',
but any information might already be rooted in later legends.

The Flood marks a central element for the first dynasty of
kings/emperors.
Manifold efforts were made to place it at the start of the Chinese
chronology.
Some have given specific years, while others consider the problem
intractable (Table \ref{tab:xiadynastie}).

\subsection{Archaeological evidence}

An archaeological historicity can only be verified for the second
dynasty, the Shang, which started in the mid-second millennium BCE.
There are indications that Xia and Shang were coexisting and
interacting spheres of influence vying for supremacy
\cite{chi_fairbank, chi_pang_1987}.
Even some coeval intervals in the king lineages of the two dynasties
have aroused suspicion.

A long-standing debate is related to whether or not the Xia was
identical with the so-called Erlitou culture, an urban society,
that existed in the Yellow River valley at the same time.
Guided by some geography texts, excavations were performed at a
site Yangcheng in Honan.
The place would be consistent with Anyi, the probable residence of
the Xia kings, though there are different locations surmised.
Pottery was recovered, ceramic vessels, and pieces of bronze.
One sample has a tree-ring-corrected age of 1900$\pm$70 BCE
\cite{chi_pang_1987}.
However, this gives no proof for the Xia, since the affiliation of
the archaeological samples remains unclear.
Because of this deficiency, the time for this dynasty is given
very, very roughly depending on the method of analysis, see Table
\ref{tab:xiadynastie}.
Its starting point may fall at some time within the interval
between 2200 and 1900 BCE.
The endpoint of the Xia is not known for the same reasons.

Seventeen monarchs are known from the Xia, but the lengths of their
regencies are rather estimated.
The ancient historian Sima Qian (ca.\ 145--86 BCE) gave a relative
order for their sequence,
but even the names have been assigned posthumously, for they are
not well identified.
Most events from those times originate from texts compiled during
the Eastern Zhou Dynasty (from 771 BCE onwards).
How they were transmitted over many centuries remains unknown, too.
Actually, the whole existence of the Xia rests on legendary stories.

\subsection{Agglomeration of planets}

The close approach of three or more planets is usually a
magnificent, eye-catching sight in the sky.
In the history of mankind there were social turmoils at such
occasions.
This issue is widely ramified and has been discussed extensively
under astronomical as well as historical aspects elsewhere.

Almost every culture considered tight configurations of planets
as the moment of creation.
It was linked to the beginning of time and set equal to the onset
of the particular calendar.
In Babylonia, India, Greece, and in the Mayan Empire astronomers
computed great cycles for the cosmos based on the re-occurrence of
special planetary meetings.
First an apocalypse was prophesied, often combined with a deluge,
thereafter a re-creation with a ``new era''.
Ancient scholars believed that \emph{any} cosmic cycle would have
started from a certain point in the past:
the day, the lunation, the year, and the course of the planets.

%%%%% Figure: Agglomeration in 1953 BCE
%
\begin{figure*}[t]
\centering
\includegraphics[width=\linewidth]{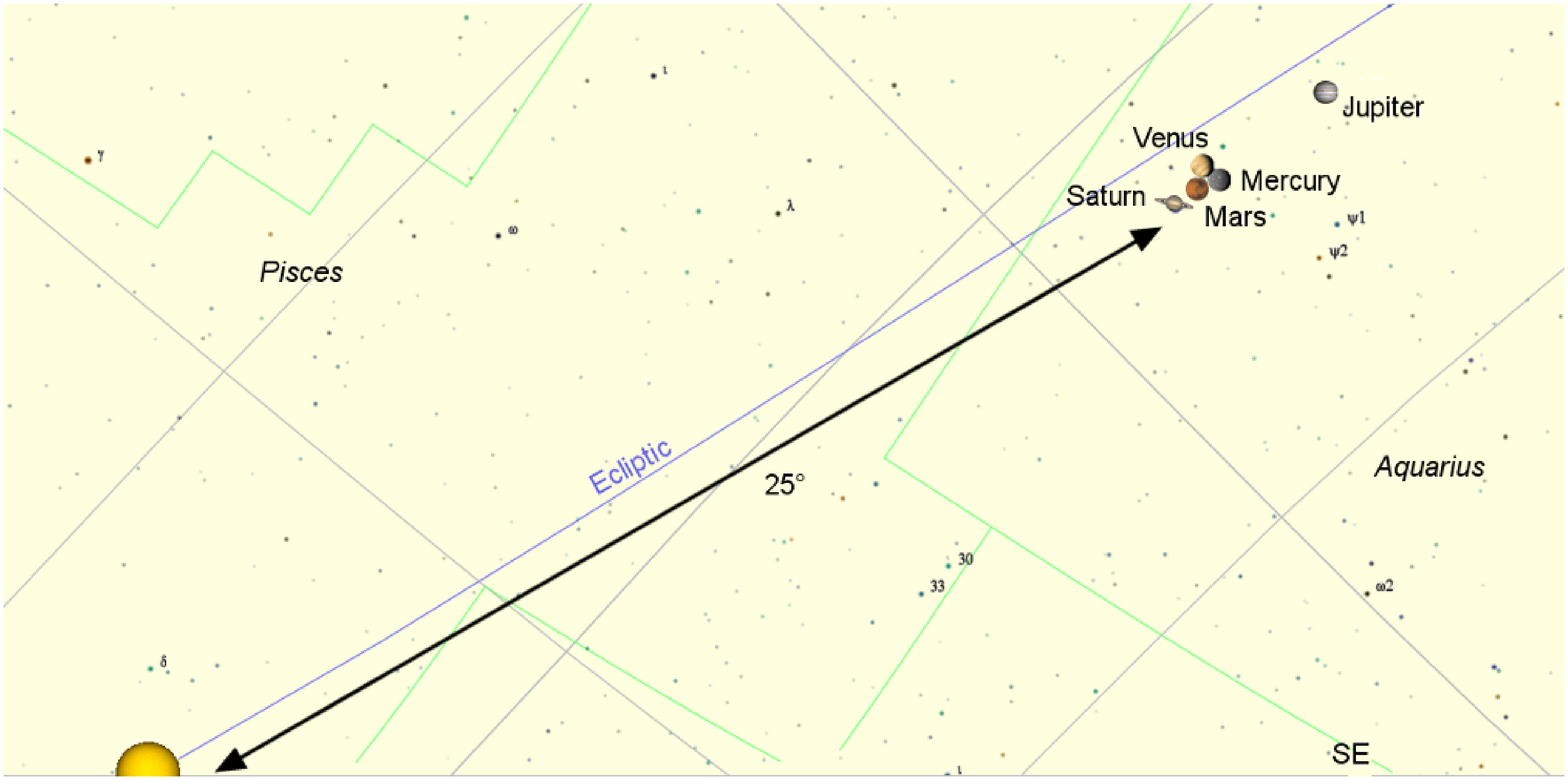}
\caption{Closest-ever agglomeration of the naked-eye planets on
    25 February 1953 BCE, 23:36 UT, at sunrise in Anyi.}
\label{fig:china1953bc}
\end{figure*}

An overwhelming clustering of planets did happen in spring of
1953 BCE.
All five classical planets gathered at dawn with the moon and the
sun joining the parade.
This would have triggered the Chinese calender.
The extraordinary grouping is accompanied by a remark in the
work by the librarian Liu Xiang (77--6 BCE)
\cite{pang-bangert}:
\begin{quote}
The original \textit{Zhuanxu} calendar began \dots\
on cyclic day 6, mon 51, year 51 (modulo 60)
at the start of spring when the sun, moon and 5 planets met
at \textit{Yingshi}, 5$^{\circ}$.
\end{quote}

The date is correlated with 5 March 1953 BCE.
The positions of the planets reached their smallest separation
seven days prior to that, namely on 26 February, with a minimum
distance of 4.3$^{\circ}$ (Figure \ref{fig:china1953bc}).
That value was never beat in the reproducible history such that
it is regarded as the narrowest grouping ever.
Robert Weitzel was the first to draw attention to the scene as
an astronomical curiosity, but he was unaware of the historical
record by the Chinese \cite{weitzel_1945}.
On 2 March, the planets were visited by the waning moon, and a few
days later the new month began with the lunar crescent in the
evening time.

The massing of planets took place in the constellation of
``Yingshi'', indeed:
It is allotted to Pegasus near the border of Pisces and Aquarius
not far from the equinox.
Pegasus was rising, but the planets could not be seen well
due to the proximity of the sun (too small difference in height).

Sometimes the above quotation is assigned to the mythological
proto-Emperor Zhuanxu who reigned two positions before Yao.
Texts depict Zhuanxu as a god who separated Heaven from Earth
and arranged the positions of the sun and stars.
At the age of twenty he became sovereign, going on to rule for
fabulous 78 years similar to his predecessors and successors.
Various authors place Zhuanxu somewhere at $\approx$2500 to
2400 BCE \cite{chi_nivison-pang}.
The ``conjunction'' of five planets on the same day would have
happened in 2513 BCE \cite{chi_brown}.
Unfortunately, there was no such clustering, even less it would
have been as conspicuous as the one of 1953 BCE.
From the historical point of view, we just may wonder why no
Sumerian reference exists to this, for their deeds are more
numerous for the same period of time.

It seems incomprehensible whether and how the knowledge about that
particular agglomeration would have survived all the centuries
down to its documentation in the 1st century BCE.
It is most likely that the event was computed backwards and then
defined to be the starting point of the Chinese calendar.
However, there seems no appearance of astronomers having had the
skill to do so, especially, such far back in time.
It was not before the royal astronomer and mathematician
Guo Shoujing (1231--1316) realised, around 1280 AD, that the
planetary cycles are incommensurable \cite{pang-bangert}.
They could never have had a common start.
This celestial incident offers by no means a reliable framework
neither for the dynasty of Xia nor for the calendar epoch nor
for the flood.

Another concentration of planets happened at the end of December
1576 BCE, when four of the five planets assembled in Sagittarius.
Except Venus, the others convened within a circumference of
$5^{\circ}$.
Historically, the Xia dynasty was replaced by the second,
the Shang.
The new ruler claimed this clustering as his ``legitimation''
for the command.
At that occasion the calendar was reformed deploying a
60-day-cycle.
This cycle gives a special designation to each day.
The mechanism is up and running strictly and continuously to our
days without any disruptions.

Then, in late May 1059 BCE, the naked-eye planets appeared for a
rendezvous (6.5$^{\circ}$) once more, and the freshly established
Zhou dynasty declared it to its ``Heavenly Mandate''
\cite{pankenier_1980}.
Again, the calendar was manipulated and then suspended by a later
dynasty.
The Han, that just conquered power in 206 BCE, made use of another
gathering of planets in May 205 BCE, though it was less impressive
($\approx21^{\circ}$).

All kinds of calendar reforms have been, at any time, an appropriate
remedy for practising political control.
Many sovereigns tried to customise the history of the country to
their own needs.
All these ``conjunctions'' can hardly be used for dating purposes,
for they have been reckoned in later times to suit the astrological
influence ``justifying'' their policy \cite{chi_keenan_2002}.

\subsection{Xia-Shang-Zhou Project}

On the whole, the Chinese history consists of many divisions as
well as rivalling monarchs.
The times of their rulership were differently reckoned depending on
the source of information.
Their flourishing is of great importance for both historians and
astronomers.

For the purpose of a binding chronology the so-called
Xia-Shang-Zhou Project was born in 1996 \cite{chi_liu_2002}.
It collated more than 200 scientists from various fields of work
among them archaeologists, astronomers, historians, and
palaeographs.
The objective was to explore the period of action for the first
three dynasties.
The experts should reconcile fixing points for the old data with
our modern chronology.
The Project was carried out by 44 working groups, twelve of which
used mainly astronomical evidence.
As for the Xia, a decision was made with the commencement of this
dynasty placed at approximately 2070 BCE
(Table \ref{tab:xiadynastie}).

The Western world looks more critical at this national project.
Sinologists consider many conclusions as inadequate, see e.g.\
\cite{chi_keenan_2002}.
New evidence has it that the new dating by the Chronology Project
also turns out to be flawed.
Recent archaeological discoveries like the sculptural bronze
artifacts from the Meixian County undermine almost every one its
dates \cite{chi_shaughnessy_2009}.
So, the controversy surrounding the solar eclipse of our interest
during the Xia dynasty is not settled yet.

%%%%%%%%%%%%%%%%%%%%%%%%%%%%%%%%%%%%%%%%%%%%%%%%%%%%%%%%%%%%

\section{The four versions of the eclipse}

The next reference after Yao to an astronomical incident appears
in the \textit{Shujing} at the time of the fourth Xia-Emperor
Zhong Kang.
In his first year occurred that solar eclipse that became so
popular among the legends.
We quote four versions.

\subsection{Shujing}

In the order of progression, the preface to the Book III of the
\textit{Shujing} points to ``The Punitive Expedition of Yin''.
The prince of Yin was sent out to punish He and Ho because they
allowed the days to get into confusion \cite[p3]{chi_legge}:
\begin{quote}
(1) When Zhong Kang commenced his reign over all within the four seas,
the prince of Yin was commissioned to take charge of the imperial
armies.
At this time He and Ho had neglected the duties of their office,
and were sunk in their private cities, and the prince of Yin
received the imperial charge to go and punish them.
\end{quote}

Then, Book IV describes the mission of the prince on behalf of the
emperor in more detail.
In the verses 2 and 3 the prince makes a military announcement.
He addresses his forces on the objective of the expedition and says
that the stability shall be restored in the land.
So far, there is no mention of an eclipse.
The reader does not come to know who the two persons were, nor
their duties, nor how they committed their crime.
The laws are put forward and then it is said \cite[p162]{chi_legge}:
\begin{quote}
(4) Now here are He and Ho.
They have entirely subverted their virtue, and are sunk and lost
in wine.
They have violated the duties of their office, and left their posts.
They have been the first to allow the regulations of heaven to get
into disorder, putting far from them their proper business.
On the first day of the last month of autumn, the sun and the moon
did not meet harmoniously in \textit{Fang}.
The blind musicians beat their drums;
the inferior officers and common people bustled and ran about.
He and Ho, however, as if they were mere personators of the
dead in their offices, heard nothing and knew nothing; ---
so stupidly they went astray from their duty in the matter of
the heavenly appearances, and rendering themselves liable to
the death appointed by the former kings.
The statutes of government say,
``When they anticipate the time, let them be put to death without
mercy;
when they are behind the time, let them be put to death without
mercy.''
\end{quote}

The concluding three verses of the Book IV contain invocations to
the enemies to surrender, since the legitimation for the punishment
was ordered by the heavens. ---
The reader is informed about a turmoil in the country.

\subsection{Bamboo account}

The second passage is found in the \textit{Annals of the Bamboo Books}.
The incident is recorded as follows \cite[p119]{chi_legge}:
\begin{quote}
(1) In his first year, which was \textit{ke-ch'wo} (26th of the
year cycle = 1951 BCE), when the emperor came to the throne,
he dwelt in \textit{Chin-sin}.
(2) In his 5th year in the autumn, in the 9th month, on the day
\textit{kang-siu} (47th of the day cycle), which was the first day
of the month, there was an eclipse of the Sun, when he ordered the
prince of Yin to lead the imperial forces to punish He and Ho.
\end{quote}

The calendric and territorial information will be analysed in the
next section.
About the capital, which is said to be \textit{Chin-sin},
the translator, James Legge \cite{chi_legge}, remarks in a footnote:
``The site of \textit{Chin-sin} is not well ascertained.
The dictionary places it in the district of Wei, department of
Shan-tung.
Others --- more correctly, I think, --- refer it to the district
of Kung, department of Ho-nan.''

Deviant from the \textit{Shujing} it is mentioned here that the
eclipse occurred in the fifth year of the Emperor.
When reading the text independently, the crime of He and Ho
becomes not clear.
It is not explained what the eclipse has to do with them.
Their fault could lay some time before it.
In other words:
there are two people being punished on a day when, by chance, a
solar eclipse happened?
\cite{chi_fotheringham_1921}

Moreover, the name of the punishing prince is not given, again.
``Yin'' is --- as in the \textit{Shujing} above --- the region
where he lived or descended from.
The modern encyclopaedia give a mountains to the North of China.

\subsection{Chunqiu}

The confusion widens as one takes notice of the third text by
Zuo Qiuming.
He writes in his commentary to the ``Spring and Autumn Annals'',
\textit{Chunqiu}, about an eclipse that was to take place in the
6th month, in summer.
At the decisive point a discussion is going on between a priest and
a historian.
The historian can be set equal to the astronomer or astrologer who
takes care of the chronology and calendar.
The two converse about the rites to be performed in the case of
the eclipse:
to beat the drums, wearing of clothes, and presenting gifts.
While quarrelling over what protocol should be followed, the
historian/astronomer intercalates and recalls a former eclipse
long ago \cite{chi_fotheringham_1933, chi_wang-siscoe}:
\begin{quote}
The Sun already passed the equinox but has not arrived at the
solstice.
When any calamity happens to the three celestial bodies
(i.e.\ Sun, Moon, and planets), the various officials put off
their elegant robes, the king does not have his meal (table)
fully spread, and withdraws from his principle chamber, till the
time is past. (\dots )
This is what is written in the \textit{Xia Shu}:
``The Sun and Moon could not live harmoniously in their place,
the blind beat the drum, low rank officials mounted the horses,
and people ran up in haste.''
That is said of the first day of this month --- it was in the 4th
month of Xia, which is called the first month of summer.
\end{quote}

The words within the quotation marks used by the royal astronomer
are identical with those in the \textit{Shujing}.
There seems no doubt that they refer to the same solar eclipse
of Xia.
According to Schlegel \& K\"uhnert,
the word ``equinox'' alludes to the spring, and the instructions
would only apply to eclipses in summer \cite{chi_schlegel-kuehnert}.
The circumstances would resemble those old days of Xia.

Though not dealing directly with the story of He and Ho, the text
throws instructive light on the eclipse customs in ancient China.
Finally, we get the impression that the aftermath of the solar
eclipse encompassing He and Ho remained in mind for very long.
Now we have three quotations at hand bearing reference to the same
incident, but they are ambiguous.

\subsection{A modern version}

In modern times, a completely different variant emerged.
The misdoing of the two astronomers would be contrary:
they foretold a solar eclipse that did not occur.
The concept is based on the idea that the monarch possessed the
power of darkening the sun whenever he wanted.
Actually, he relied on a man skilled in foretelling eclipses.
The two astronomers got secret information about the next date by
a subterfuge and wanted to benefit from it over the emperor.
The emperor gathered the public upon the prognostication in order
to distinguish himself by the miracle of commanding the sun.
But the foreteller intentionally included a wrong date, and the
occultation failed to appear.
After the proclamation had gone wrong, the monarch was embarrassed
so much that the doom of the sinners was the same.

This alternative entertains with a love story and intrigues.
It has its roots in the novel ``Sonne, Mond und Sterne''
by A.G.\ Miller, which is a pseudonym of
the German-Austrian writers Marie Louise Fischer (1922--2005)
and Hans Gustl Kernmayr (1900--1977).
Although the book is categorised as non-fictional, the storyline
suits rather a thrilling TV movie than sound history
\cite{chi_herrmann}.
This version can be rejected, for in those times there was
definitely no-one to presage a solar eclipse.

Within the alleged $\approx$180 years between the first Emperor Yu
and the fourth Zhong Kang about 15 solar eclipses could have been
visible, mostly partial ones.
This is much to little to discover an astronomical cycle, even if
all of them would have been observed \cite{chi_schlegel-kuehnert}.
Another constraint concerns the fact that there was no eye
protection for monitoring the sun permanently.
It was not useful, anyway, and thus only those eclipses would be
seen that are almost total, or close to the horizon at sunrise
and sunset.
Many partial obscurations would have passed unnoticed.
Therefore, the two unfortunate astronomers could never have met
the requirements of their office.

%%%%%%%%%%%%%%%%%%%%%%%%%%%%%%%%%%%%%%%%%%%%%%%%%%%%%%%%%%%%

\section{Analyses and interpretations}

Countless attempts were made to identify the mysterious eclipse in
spite of objections by historians.
From the first text, the \textit{Shujing}, only one line has an
astronomical relevance:
The Sun and the Moon did not meet harmoniously in \textit{Fang} on
the first day of the last month in autumn.

One might think that it will be simple to find a suitable event
having the location and date at hand, but, unfortunately, it is not.
The main difficulty concerns the interpretation of any information:
the specification of the season is disputed as well as the
celestial meaning of ``Fang''.

\subsection{Constellation ``Fang''}

Usually, \textit{Fang} is assigned to an area of about 5$^{\circ}$
between $\beta$, $\delta$, $\pi$ and $\sigma$ Scorpii,
a region in the ``pincer'' westward of Antares.
A few stars from Libra and Ophiuchus would belong to it, too
\cite{chi_williams_1863}.
The Sun needs 5 days to cross this part of the sky.
According to the Xia calendar, it is the ninth month, the month
after that which contains the autumnal equinox.

Other commentators consider \textit{Fang} as undetermined but
understand it more general as a ``planetary mansion'' or
``domicile'', i.e.\ those 28 Chinese lodges along the ecliptic
which can be visited by the Moon
\cite{chi_schlegel-kuehnert, chi_chalmers}.
Each of the constellations has a different extent:
the smallest about 1$^{\circ}$, the largest 31$^{\circ}$.
Thus, \textit{Fang} would not be used as a fixing point for the
position of the sun.
Also, a calligraphic confusion could not be excluded among the
thousands of symbols in the Chinese language.

The statement that the sun and the moon ``did not meet
harmoniously'' also turns out quite ambivalent.
Such an expression exists nowhere else in Chinese literature
except for the \textit{Shujing}.
When denoting an eclipse, the common usage was something like
``the sun was devoured'' or ``it lost its light''
\cite{chi_schlegel-kuehnert}.
A variant of the original phrase might read ``the sun and the moon
could not live peacefully together in the sky''
\cite{chi_wang-siscoe}.
Another translation is ``[they] were not in agreement''
\cite{chi_williams_1863}.
The crucial point applies to the Chinese symbol for the sun:
Depending on the context it could be altered in ``time'' or
``celestial body'', or it can denote a morning hour between 7 and
9 a.m.\ \cite{chi_oppolzer_1880}.
This gives a different meaning to the whole line.
Since most experts base their interpretation on the authority
of James Legge \cite{chi_legge}, the author of this paper will
stick to that.
It is generally agreed upon that we are treating a solar eclipse
here.
The chaos among people and the consequences for the victims suggest
that no other phenomenon is the subject.
In addition, the debate in the \textit{Chunqiu} explicitly deals
with this issue.

As far as the instant of the eclipse is concerned, the last month
of autumn is specified (\textit{Shujing}), or the first month
of summer (\textit{Chunqiu}), or the whole event could have been
retrieved by computation very much later.
In historiography such acts were carried out quite often, as many
examples show like the so-called ``Star of Bethlehem'' or birth
dates of celebrities with political power.
%%Beispiele: Kali-Yuga, Bethlehemstern, Mohammed
Some historians prefer to fix the eclipse exactly to the day of
the autumn equinox.
In fact, it was lying in the constellation of \textit{Fang} about
2150 BCE, and it would correspond to 11 October of our calendar
\cite{chi_stockwell_1895}.

Keeping in mind that the books were written in the 5th century BCE,
one would have to consider the luni-solar precession when
calculating backwards.
The Chinese were not aware of the precession at the time of
Confucius, though.
It was discovered in the 4th century AD, more than 500 years after
Hipparchus \cite{chi_russell_1895}.
Hence, there is a suspicion of a deliberate modification by the
restorers of the \textit{Shujing} as such as they substituted
``autumn'' for the reading ``winter'' because \textit{Fang} had
moved to that season.
Or, taking the other perspective, autumn would be the original and
\textit{Fang} was emended for agreeing with the lifetime of the
writer.
Most present-day analysts, however, give credence to the last
month of autumn, though there is no reason.
The question about the season must remain unsettled.

%%%%% Table: Xia capitals
%
\begin{table*}[t]
\caption{Suggested capitals for the Xia dynasty.}
\label{tab:xiacapitals}
\centering
\begin{tabular}{llccc}
\hline
\rowcolor{grey20}
Place      & Today's name  & Latitude & Longitude & Reference \\
\hline
An Yi Hsien& Anyi      & $35^{\circ}\; 5^{\prime}$ N  & $110^{\circ} 58^{\prime}$ E &
           \cite{chi_oppolzer_1880, chi_schlegel-kuehnert, chi_russell_1895} \\
Tay-kang-kien& Taikang & $34^{\circ}\; 7^{\prime}$ N  & $114^{\circ} 57^{\prime}$ E &
           \cite{chi_rothman_1840, chi_russell_1895} \\
Tschin-sin & (in Honan) & $34^{\circ} 54^{\prime}$ N & $114^{\circ}\;\, 6^{\prime}$ E &
            \cite{chi_schlegel-kuehnert} \\
Tschin-sin &(in Shan-tung)& $36^{\circ} 46^{\prime}$ N & $119^{\circ} 20^{\prime}$ E &
            \cite{chi_oppolzer_1880} \\
Yangchen   & Dengfeng  & $34^{\circ} 27^{\prime}$ N & $113^{\circ}\;\, 1^{\prime}$ E &
            \cite{chi_pang_1987} \\
Yanshi     & Luoyang   & $34^{\circ} 40^{\prime}$ N & $112^{\circ} 27^{\prime}$ E &
             \cite{chi_fairbank, chi_henriksson_2008} \\
\end{tabular}
\end{table*}

\subsection{Cyclic day}

The second quotation from the \textit{Bamboo Books} provides an
exact day in the shape of a cyclic name:
\textit{kang-siu} (\#47).
It is generated by combining one of ten ``heavenly stems'' with
one of twelve ``earthly branches''.
The 60-day-cycle acts like an analogy to our 7-day-week:
It runs without an interruption eversince.
Admittedly, it was introduced in the second dynasty around 1500 BCE
--- for the Xia meaningless.
If still trying to analyse it, discrepancies arise being at odds
with the eclipse dates.

Richard Rothman (1800--1856) approved of the cyclic day name
corresponding to the eclipse of 13 October 2128 BCE
\cite{chi_rothman_1840}.
The eclipse would have had a magnitude of 0.875 at noonday.
The historian of astronomy John Fotheringham (1874--1936) referred
to the \textit{Bamboo Books} giving 28 October 1948 BCE as the
correct day for the name --- 3$\times$60 years later than Rothman's
result \cite{chi_fotheringham_1921}.
Unfortunately, there was no eclipse that day, not even a new moon.
Fotheringham put forward that the eclipse did not match the lifetime
of Zhong Kang and his reign would have to be shifted within a
century or two one side or the other of 2000 BCE.
Instead, he emphasized the clearness with which the
\textit{Shujing} makes the offence of the astronomers' neglect
of the calendar.
It went out of order, and this became evident through the
occurrence of the eclipse on an unexpected day,
i.e.\ not on the first day of the month as it should be.
This drew attention to their negligence and convicted them of the
error of issuing a wrong calendar \cite{chi_fotheringham_1933}.
They were executed for the false reckoning and not for the eclipse.
Finally, Fotheringham concluded that the record did not permit an
identification of the date at all.

Before that, the historian John Williams (1797--1874) used the
name of the cyclic year in the previous verse, and obtained 2158 BCE
as the year of the accession of the emperor
\cite{chi_williams_1863}.
But the cycle of names was only applied to years during the Han
dynasty from $\approx$200 BCE, after the completion of the
\textit{Bamboo Books}.
When looking into various encyclopaedia, the durations of rulership
for all emperors prove contradictory.
The information about the cyclic names, whether related to the day
or year, must be a later insertion based on a doubtful backreckoning
\cite{chi_pang_1987}.

\subsection{The capital of Xia}

In addition to the arguments above there are further disharmonies:
The location of the residency of the Xia dynasty is not well known.
It changed permanently with the emperor, and could be situated in
Anyi or Taikang, which is 500 km farther to the East
(Table \ref{tab:xiacapitals}).

Another problem is that the text gives no hints neither to the
time of day nor to the magnitude of the eclipse:
total or a very high degree of obscuration.
For example, there are no accompanying phenomena mentioned like a
sudden darkness or the visibility of stars or allusions to the
corona.
If the eclipse was not total, the search would expand to numerous
partial eclipses.

Also, it is not said what space of time passed after the eclipse
when the ``Punitive Expedition'' against the astronomers was
set off.
The punishment could rest upon political motives.
In the context of history a war against an usurper followed
lasting for two years \cite{chi_fotheringham_1921}.
The procedure of punishment is not told, either.
A ``decapitation'', as widely depicted, is not stated.
The victims could be hanged or stoned to death or be ``punished''
in any other way.
The reader of the text is even not informed whether the punishment
was actually inflicted or not.
It seems enough to have it declared that they merited death.

Some modern authors stretch out the interpretation as much as
they do not regard the names He and Ho as individuals but rather
the titles of officers because they are called this way under
the pre-emperor Yao \cite{chi_fotheringham_1933}.
William Whiston (1667--1752), a theologian and mathematician who
succeeded his mentor Isaac Newton (1643--1727) at the University
of Cambridge, believed that both Emperors Yao (counting \#7 in the
succession series) and Zhong Kang (\#12) were nearly contemporary
to one another \cite{whiston_1734}.
Again others suggest that it was only \emph{one} astronomer named
He-Ho \cite{chi_wang-siscoe}.
He would be the usurper of the rebellious tribe and assembled
a considerable military force to overthrow the emperor.
This discomfited Zhong Kang, so he took advantage of the eclipse
to attack He-Ho.

All these imponderables raise the thought that the subject in
the \textit{Shujing} is not the eclipse itself.
The account does not focus on the astronomical event but seems
to give a moral sermon for deterrence.

%%%%% Table: Suggested Dates
%
\begin{table*}[t]
\centering
\caption{Dates of proposed solar eclipses related to He and Ho.}
\label{tab:xia-sofi}
\begin{tabular}{lcll}
\hline
\rowcolor{grey20}
Date [BCE] &Type& Commentator (publ.\ year)  & Reference \\
\hline
2128, Oct 13 & P & Yi Xing ($\approx$700)    & \cite{chi_liu-etal_2003} \\
2128, Oct 13 & P & Guo Shoujing ($\approx$1280) & \cite{chi_brown, chi_liu-etal_2003} \\
2155, Oct 11/12&A& Antoine Gaubil (1732)     & \cite{gaubil} \\
%                   % auch: {chi_rothman_1840} ; {chi_qihan} ; {chi_brown}
2137, Oct 22 & A & William Whiston (1734)    & \cite{whiston_1734} \\
2007, Oct 24 & T & Nicolas Fr\'eret (1796)   & \cite{chi_oppolzer_1880, chi_schlegel-kuehnert} \\
2128, Oct 13 & P & Richard Rothman (1837)    & \cite{chi_rothman_1840}\\
2156, Oct 22 & T & Johannes von Gumpach (1853) & \cite{chi_oppolzer_1880}\\
2128, Oct 12 & P & John Chalmers (1861)      & \cite{chi_chalmers}\\
\multicolumn{2}{l}{(2158 -- accession)} & John Williams (1863) & \cite{chi_williams_1863}\\
2137, Oct 22 & A & Theodor von Oppolzer (1880) & \cite{chi_oppolzer_1880}\\
2165, May  07  & A & Schlegel \& K\"uhnert (1889) & \cite{chi_schlegel-kuehnert}\\
(1905, May  12) & T & Schlegel \& K\"uhnert (1889) & \cite{chi_schlegel-kuehnert} \\
2136, Oct 10 & P & John N.\ Stockwell (1895) & \cite{chi_stockwell_1895}\\
2137, Oct 22 & A & Samuel M.\ Russell (1895) & \cite{chi_russell_1895} \\
(1948, Oct 28) & --- & John N.\ Fotheringham (1921) & \cite{chi_fotheringham_1921, chi_brown}\\
2159, Jun  29 & T & F.\ Crawford Brown (1931)& \cite{chi_brown} \\
2110, Oct 23 & P & Liu Chao-Yang (1945)      & \cite{chi_wang-siscoe}\\
1876, Oct 16 & A & Kevin D.\ Pang (1987)     & \cite{chi_pang_1987}\\
1912, Sep 24 & A & Kuniji Saito (1992)       & \cite{chi_liu-etal_2003}\\
1961, Oct 26 & T & G\"oran Henriksson (2008) & \cite{chi_henriksson_2008}\\
1903, Sep 15 & H & {\it this work, Sec.\ \ref{ek-xia}} (2020) & \cite{finsternisbuch}\\
\end{tabular}
\end{table*}

%%%%% Figure: Eclipse Paths
\begin{figure*}[t]
\includegraphics[width=\linewidth]{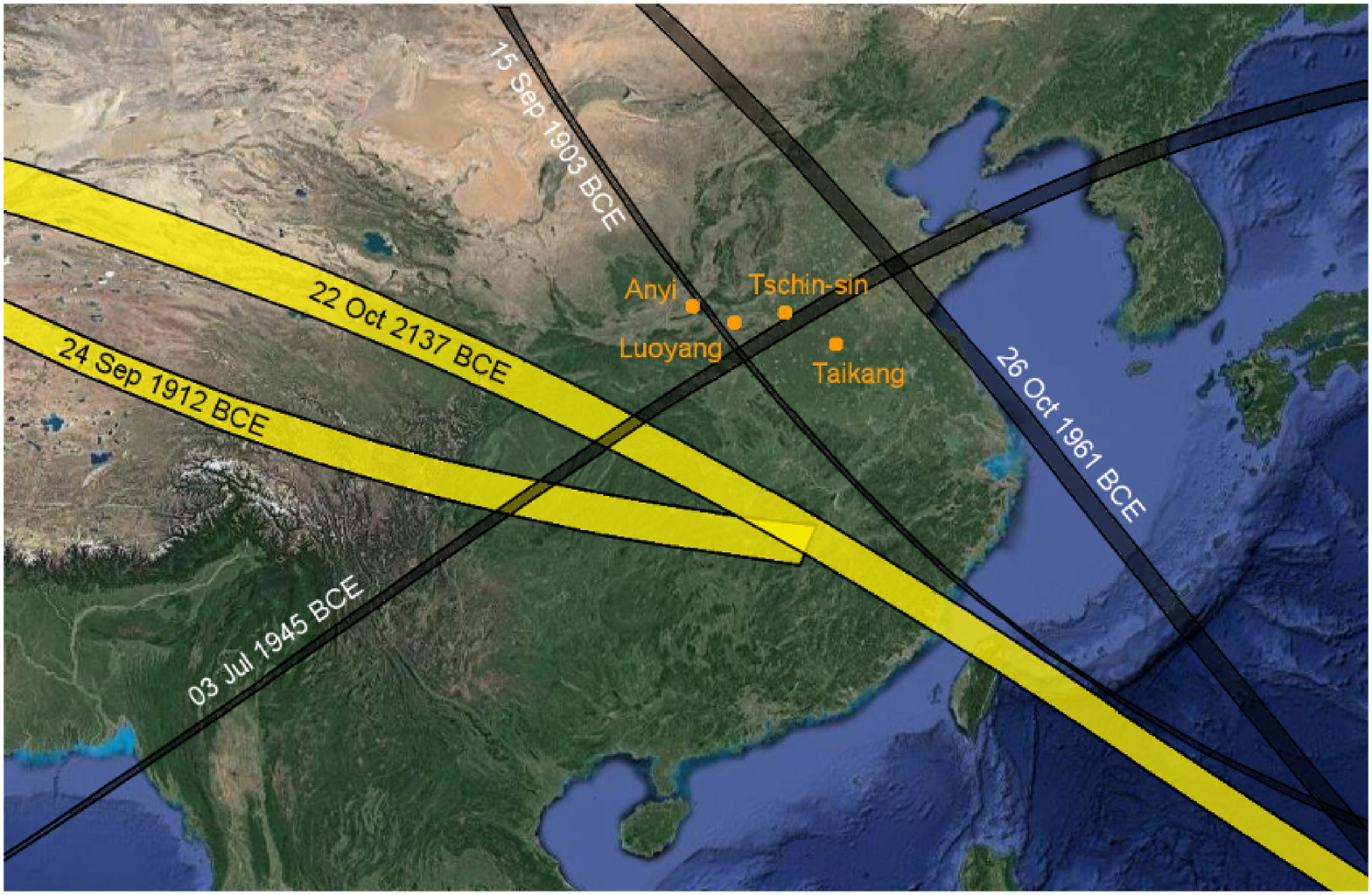}
\caption{Central zones of two annular eclipses (yellow) and tree
    total eclipses (grey), one of which is of 1903 BCE.}
\label{fig:hihoeclipse}
\end{figure*}

%%%%%%%%%%%%%%%%%%%%%%%%%%%%%%%%%%%%%%%%%%%%%%%%%%%%%%%%%%%%

\section{History of suggested eclipses}

The search for a solar eclipse falls into an interval of about
300 years depending on the assumed commencement of the Xia dynasty.
Table \ref{tab:xia-sofi} lists a selection of suggested dates
sorted by the year of publication.

%%%%% first proposals of the early 19th century
%
The findings scatter tremendously.
The first to introduce the Chinese works to the Western world was
the Jesuit Antoine Gaubil (1689--1759).
He rejected the year of 2128 BCE from medieval times, by the help
of a Chinese assistant, because the eclipse would not have taken
place in Beijing \cite{gaubil}.
Instead, he introduced the date of 12 October 2155 BCE.

It was William Whiston who first referred to Edmond Halley's
discovery, in 1695, of the ``acceleration of the Moon's mean
motion''.
In 1734 Whiston pointed out the small quantity of obscuration
($<$1/12th) of Gaubil's proposal, which would ill agree with the
capital punishment of two astronomers for not taken notice of so
small an eclipse as this was \cite{whiston_1734}.
He replaced it by one later Saros to appear more striking in many
parts of China.
These early computations from the 18th century should be taken with
some caution, for the method of determination was not technically
mature:
neither the deceleration of the earth's spin (clock error
$\Delta T$) nor the Besselian Elements were worked out.
However, the date by Gaubil still prevailed for long.

An improved value for $\Delta T$ was included by Charles-Louis
Largeteau (1791--1857).
In 1840 he re-calculated the ephemeris of the moon and found that
Gaubil's eclipse of 2155 BCE occurred when China resided on the
night side \cite{chi_schlegel-kuehnert}.
Also, the year 2128 BCE will have to be definitely excluded,
in spite of a revival by John Chalmers in 1861, who was unaware
of the work by Largeteau \cite{chi_chalmers}.

%%%%% Russell + Oppolzer (2137)
%
In our time, the date of 22 October 2137 BCE is mentioned most
often, as suggested by Whiston (Figure \ref{fig:hihoeclipse}).
The case was examined by the astronomer and mathematician Samuel
Russell (1856--1917) in detail and seemingly discussed in the most
convincing way \cite{chi_russell_1895}.
Theodor Ritter von Oppolzer (1841--1886), the time-honoured
authority on eclipses, reached the same conclusion a decade earlier
\cite{chi_oppolzer_1880}.
Popular science repeats this date constantly without elucidating
the origin of the information.

However, that eclipse of 2137 BCE was not total but annular,
and the magnitude in Anyi was about 0.875.
A complete darkness was never achieved at any moment.
Moreover, the date does not correspond to the (speculative)
cycling day \#47 from the \textit{Bamboo Annals}.
The cyclic name would perfectly coincide with the former 13
October 2128, which must be excluded in any case.

In the year following 2137, there was another eclipse also visible
in the northern hemisphere: 11 October 2136 BCE.
Russell gives the time of onset at 4:30 p.m., and the magnitude
in Anyi was about 0.58 \cite{chi_russell_1895}.

%%%%% Kevin Pang (1876)
%
Including the concept of the Great Flood, which was the cause for
the accession of the first Emperor Yu and put by archaeologists at
approximately 1925 BCE, then the late date by Kevin Pang gains a
surprising attention \cite{chi_pang_1987}.
It could reduce some conflicts from the historical point of view,
though not eliminate them from the astronomical.
The agglomeration of planets cannot be integrated, for example,
into Yao's or Yu's lifetime and his successors.
And, again:
the eclipse of 1876 BCE was annular, and the central track passed
over Siberia much to the North.
In Anyi, only a small partial coverage (mag = 0.386) would be
observed shortly after sunrise at 8:30 a.m.\ local time.

%%%%% G\"oran Henriksson (1961)
%
The archaeoastronomer G\"oran Henriksson
argues in favour of the eclipse of 26 October 1961 BCE because it
was total and crossed the territory of Yin where the punishing
prince had his residence \cite{chi_henriksson_2008}.
Henriksson locates the region to the Northeast of China at the
estuary of the Yellow River about 200 km South of Beijing.
In Anyi the eclipse would be seen partially with a magnitude
of 0.863.
The sun had a distance of 10$^{\circ}$ from $\pi$ Scorpii, the
central star in \textit{Fang}.

%%%%% Xia-Shang-Zhou-Project (4 dates)
%
After providing new limits for the Xia dynasty by the
Xia-Shang-Zhou Project, none of the dates suited for the eclipse
except Henriksson's.
A plenty of new candidates emerged:
2043, 2019, and 1970 BCE that are not listed here
\cite{chi_liu_2002}.

%%%%%%%%%%%%%%%%%%%%%%%%%%%%%%%%%%%%%%%%%%%%%%%%%%%%%%%%%%%%

\section{The hybrid eclipse of 1903 BCE}
\label{ek-xia}

%%%%% Eigener Vorschlag: -1902
%
Most commentators commit themselves to the constellation of
\textit{Fang}.
They examine ``automatically'' a date in October.
If one is willing to look at the putative course of events with
some reservation, he comes across an eclipse less considered:
15 September 1903 BCE (Figure \ref{fig:hihoeclipse}).

The hybrid solar eclipse flashed by in a distance of 20 km from
Anyi.
The width of totality was as narrow as 15 km.
An observer would experience a full obscuration of the solar disk
for about 12 to 15 seconds, if he stood on the central line.
The rotation rate of the Earth has been varying to such an extent
in the past that it must be taken into account whether a solar
eclipse could have been visible from a certain location or not.

The Earth's rotation is an exhaustive issue out of the scope of
this paper --- see the many works by Richard Stephenson (e.g.\
\cite{stephenson-morrison_2005}) or \cite{finsternisbuch} for the
most recent overview.
The clock error, termed $\Delta T$, is defined as the difference
between the uniform timescale, based on the celestial motion of the
sun and moon (Ephemeris Time, ET), and our standardised civil
timescale which is used for the rotation of the earth (Universal
Time, UT).
For the remote past, that difference is modelled by a parabola:

\begin{linenomath}   %%% Zeilennummern trotz math-Absatz (Begleiter, 3.5.1)
\begin{equation*}
   \Delta T = \text{ET} - \text{UT} = -20 + ct^2
\end{equation*}
\end{linenomath}
with $c \approx 32$ sec/cy$^2$ and $t$ in centuries (cy) before
1820.
This formula corresponds to a regular and systematic slow-down.
It comprises the tidal friction as well as other seasonal effects
having influence on the behaviour of the earth's rotation.
We also know that there are irregular fluctuations superimposed
on it and destroy a strict validity of the formula.
If something unexpected happens, e.g.\ an earthquake slightly
altering the moment of inertia, a systematic error enters and
accumulates over time.
It puts the backreckoning into serious trouble:
the path of the eclipse can be shifted to either side of the
mean longitude.
An extrapolation beyond 500 BCE turns out somewhat dangerous.

Although a precise localisation of the eclipse track of 1903 BCE
cannot be given as long as the account in the \textit{Shujing}
lacks of a detailed description concerning totality, the social
aftermath suggests a high magnitude of obscuration.
Even without exactly determining the belt geographically, the
extrapolated $\Delta T$ lies within the expected error bars.
Fred Espenak's map provide a $\Delta T=$ 44,058$\pm$3,328 s
(12h 14m $\pm$ 55m) \cite{chi_espenak}.
An error of 55 minutes produces a tolerance of 13,75$^{\circ}$ in
longitude.

However, the observer does not need to find himself inside the
zone of totality to become deeply touched by the sudden loss of
light thinking the end of the world is near.
% and takes him into abyss.
The ideal position will be even unlikely.
An eclipse magnitude of 0.996, as the \emph{average} value
provides, would do the job, too.
An incisive darkness, as in nighttime, in order to make stars
visible, is not required.
It rather supports the correctness of the empirical formula for
extrapolation by Stephenson \& Morrison
\cite{stephenson-morrison_2005}.

The advantage of this date is the reproduction of the historical
corner stones.
It would involve a liberate interpretation of the manuscripts.
Historians will have to disengage themselves from the long counts
of regents, and the durations of their governances must be mapped
in a more compact way.
Our new framework is outlined as follows:
\begin{itemize}
\item[1953:] Agglomeration of planets with a starting point of the
   calendar --- if this issue should matter at all;
\item[1925:] then, a Great Flood with an unsuccessful emperor
  after Yao (in accord with \cite{chi_wu-etal_2016});
\item[19**:] then, an effective remedy of unknown duration under
  the first Xia-Emperor Yu (see Table \ref{tab:xiadynastie});
\item[1903:] and, finally, the eclipse during the rulership of the
  fourth (?) Emperor Zhong Kang.
\end{itemize}

Two pre-emperors and four Xia-emperors, at least, would have to be
squeezed in a time interval of about 50 years.
Previously they were attributed much longer periods of regency.
So far, the time spans for the four Xia emperors are allocated at
45 years (Yu), 10 years (his son), 29 (grandson), and 13 years
(Zhong Kang himself), as given in the \textit{Bamboo Annals}.
The German Wikipedia gives 58, 29, 29, and 13 years, respectively,
referring to ancient Chinese historians whose information will be
outdated by now \cite{chi_w-en}.
The French version provides 8, 9, 29, and 13 years, respectively,
crediting the historian Henri Cordier (1849--1925).
More recently it was suggested that Emperor Yu's regency might
have lasted between 1914 and 1907 BCE
\cite{chi_nivison-pang}.
This absolute timeframe would still conflict the flood and should
be shifted further back in time.

That leads to the suspicion that the whole genealogy of the Xia
should also be examined anew.
Of the 17 known names (\textit{Bamboo Books}), the shortest regency
is 10 or 11 years (king \#2 or \#15 and \#16, respectively),
and the longest 59 years (king \#11).
One may ask whether there was no-one within the four centuries,
in which the dynasty lasted, holding the office for less than a
decade?
Probably some of them were too insignificant for a mention, and,
thus, they were omitted by the historians who copied, transmitted,
or restored the information.
Speculations on interregnums were also launched
\cite{chi_nivison-pang}.
Already the compression of the whole length of the dynasty would
put back the lifetimes into a more realistic view.
The ``average reckoning'', as performed by Kevin Pang
\cite{chi_pang_1987}, becomes void and seems not very scientific,
anyway.
In a time span of $\approx$20 to 25 years between the flood and
Zhong Kang, four emperors could easily find their place.

While the chronology of rulers is given back to the historians,
astronomers have to accept some other weaknesses:
The constellation \textit{Fang} does not fully agree with the
account, and the literal minuteness must be abandoned in favour of
an approximate sighting.
In September, \textit{Fang} is setting heliacally, i.e.\ one sees
its final visibility at dusk before disappearing for the conjuction
with the sun.

This raises the general question about the scene of the eclipse:
Would anyone really take care for celestial constellations while
a dramatic phenomenon with a blackened sun falls upon the
unprepared and a civil commotion is going on?
Would the observer check for some stars at which the
``non-harmonic meeting'' of the sun and moon occurred?
Or is it more likely that he guessed the time and place of the
event when the state of affairs have settled down?
And why would anyone keep such a detail for so many generations
after him? ---
We believe that the dispensable note on the constellation would
rather be sought after the routine gradually returned to everyday's
life \cite{finsternisbuch}.
It could have taken some weeks of regeneration when the
description would be put to record unless it did not slip at all.
Then it was copied without questioning by one writer after another
till it became an immovable part of the lore.
At the restoration of the \textit{Shujing}, either the season or
the constellation could have been readjusted with the calendrical
precession of the wrong era.

The cyclic day name does not match any dates, but it becomes
completely irrelevant because it was erroneously back-calculated
in medieval times when the wrong position of the sun in the sky
was assumed.
Besides that, for the ancient people, any information about time
had a much lower value than for us today.
In the rural society, someone would not hold on a certain day after,
say, half a month or so.
He or she would just memorise the ``dreadful event'' and pass it
down in dramatic stories rather than cling to the trifles of a
constellation, or hours, or the celestial circumstances;
especially, when the storyteller is illiterate as the majority of
the ``common people'' was.
Such a loose perspective on the texts leaves behind only the
location of the observer as the single anchor for the determination
of the eclipse.
Unfortunately, even this is not explicitly specified, either.

The solar eclipse of 15 September 1903 BCE is placed reasonably
compatible with the planetary grouping and the Great Flood.
It produced the greatest magnitude in Anyi between 2000 and
1850 BCE.
In view of the badly transmitted information, it fulfils several
logic criteria.

Two more eclipses deserve a short notice
(Figure \ref{fig:hihoeclipse}).
The track of totality on 3 July 1945 BCE (9:45 a.m.\ local time)
did cross the region, and its magnitude of 0.95 must have left a
noticeable effect in Anyi, provided that weather permitted.
We disregard it because of the time-lag too short since the
agglomeration of planets.
The other suggestion by Kuniji Saito for 1912 BCE gives an annular
eclipse with the central path more than 600 km farther to the South
(mag = 0.864).
This option shows a somewhat better chronology, however, the
obscuration happened during sunset at $\approx$6 p.m.\ local time.
It seems more difficult to integrate the severity of the social
behaviour at the end of the day than if people were taken by
surprise at noon.

%%%%% Kuniji Saito (1913--2003)
%%% https://www.iau.org/administration/membership/individual/4004/
%The proposal by Kuniji Saito (1913--2003), who wrote in Japanese,
%delivered also an annular eclipse.
%The sun was near the horizon when the maximum magnitude of 0.864
%occurred at an altitude of 6$^{\circ}$.
%%% http://www.eclipsewise.com/solar/SEgmapx/-1999--1900/SE-1911Sep24Agmapx.html

%%%%%%%%%%%%%%%%%%%%%%%%%%%%%%%%%%%%%%%%%%%%%%%%%%%%%%%%%%%%

\section{Critical remarks}

The most caveats were already touched in the preceding sections.
In general, accounts on eclipses in the Chinese chronicles are
both vague and sporadic before the 8th century BCE.
In many cases the texts lack a correct understanding.
In old times, people did not know how to describe such a strange
phenomenon like the disappearance of the most important celestial
body \cite{chi_liu-etal_2003}.
Such incidents did cause much panic but they were not described
in the prevalent terms.
This makes it difficult to identify an eclipse in the records.
Many events were surely observed without a message left behind,
provided that weather permitted.

Dates of any kind are scarce in the early history of China.
Usually, the name of the regent was just recorded together with a
keyword, as the example of He and Ho shows.
Occasionally, a year appeared without the name of the regent.
The earliest date that can be taken as a reliable point in the
Chinese chronology is 841 BCE,
the start of an interregnum called ``Gonghe Regency''
\cite{chi_keenan_2002}.
Beyond that date we know almost nothing about the history except
a few names of emperors and their relationships.

The second unequivocal testimony about a solar eclipse goes back
to 780 BCE.
It shows up in the ``Spring and Autumn Annals'' (\textit{Chunqiu}).
The number of accounts on eclipses rises thereafter.
So, there exists a gap of more than one thousand years to the
\textit{Shujing} eclipse.
The lack of knowledge must be ascribed to the burning of books
in 213 BCE.
According to some opinions, astronomical/astrological works were
not affected by this measure but rather those of philosophical and
historical content \cite{chi_brown}.
Again others believe that the huge gap is due to a discouragement
induced by the fate of He and Ho ---
scholars lost their motivation to show an active interest in
astronomy, and this branch of scientific occupation was avoided for
a long time.
On the other side, astronomy was of such great importance to the
emperor that it guided his political engagements.
He could not do without.

Nonetheless, there is one subtlety in the eclipse account of
the \textit{Shujing} that should baffle:
The phrase that the Sun and \emph{Moon} did meet.
The Chinese would consequently have realised that the moon
was responsible for the eclipse phenomenon.
Would this apply to those ancient times, too?
Many other civilisations were far from such an awareness, for
they thought that the moon would shine only at night.
It is our modern comprehension correlating it to an eclipse.
If one tries to gain an imagination of the cosmic matters, it will
be necessary to observe several incidents of the same kind before
the idea dawned that it was caused by the earth's companion.
When the perception once entered belief, it would be handed down
to the successive events:
the relationship between the eclipse and the moon would be mentioned
more frequent.
Hence, doubts are justified whether the text expresses the same
what we would understand as an eclipse of the sun.
If the so-called ``meeting'' was some other strange effect in
the sky, verification fails.
Therefore, the eclipse provides the only safe pillar for dating
using astronomical means.

It becomes evident that there are manifold opinions on the ominous
eclipse in the \textit{Shujing}.
They challenge the historical origin of the book itself.
It seems more important to examine the delivery of the report
throughout the ages as well as the survival of the book
\cite{chi_fotheringham_1921, chi_rothman_1840}.
In the 20th century more and more commentators pushed the eclipse
of He and Ho into the realm of legends.
A lot of stuff therein appears mystical.
But, finally, a historical core accompanies every writing.

The experts on eclipse dating are remarkably silent today when it
comes to a specification of a clear date.
Though there is a permanent scientific improvement, e.g.\ in the
physics of the earth's rotation, and precision of computation, and
methods of analysis, it all does not bail us out in this case as
long as history does not bring forward a distinct range for the
existence of the Xia dynasty.
Most of the dates in Table \ref{tab:xia-sofi} would yield an
eclipse more than 200 years before the appointed start of the
Chinese calendar.
The conflict affects any regent of that dynasty, too.
If it should ever be possible to date this \textit{Shujing} eclipse
without doubt, then one could immediately enlighten the complete
advent of China's history.

%%%%%%%%%%%%%%%%%%%%%%%%%%%%%%%%%%%%%%%%%%%%%%%%%%%%%%%%%%%%

\section{Summary}

We examined the sources on the oldest eclipse account in history
and undertook efforts to question their authenticity.
By considering other natural occurrences of mythological rank,
we tried to date the legendary solar eclipse within the context of
the Great Flood, the origin of the calendar, and the first emperors
of the first dynasty in China, the Xia.

We came across the hybrid solar eclipse of 15 September 1903 BCE.
It must have swept through the area of Anyi, the supposed capital
of the Xia.
A correction of a few minutes to the average $\Delta T =$ 44,058 s
would shift the path to either side of the capital.
Allowing for the uncertainties, it appears consistent within the
error margins for the deceleration of the Earth's rotation.
The phrase ``the Sun and Moon did not meet harmoniously in Fang''
must be interpreted more tolerantly than previously, and deploy the
heliacal setting of the stellar constellation ``Fang''.

Our arguments in favour of this eclipse are based on the matching
of the chronological order of events:
the two astronomical fix points are the gathering of planets in
1953 BCE and this eclipse 50 years later.
They form the timeframe for evaluating the pieces of historical
evidence.
This interval could harbour an earthquake causing a dramatic
inundation for several years as well as four regents, at least.
The extraordinary long rulerships of all emperors in the Xia
dynasty need to be shortened to more realistic durations.
Probably there were more unknown names that would lower the average
time of their reigns.
Moreover, doubts are discussed whether the account describes an
eclipse at all.

%%%%%%%%%%%%%%%%%%%%%%%%%%%%%%%%%%%%%%%%%%%%%%%%%%%%%%%%%%%%

\section*{Acknowledgements}

The author thanks Jeremy Shears for important comments on
improvement.
Figure \ref{fig:greatflood} is used under the licence CC-Lizenz
BY-SA.
The results are part of the Habilitation (ch.\ 10.2) submitted to
the University of Heidelberg, Germany, in February 2020
\cite{finsternisbuch}.
%Historians refused an assessment.
This article is now published ``as is'' on {\tt arXiv}.
The entire work was accomplished under severe pressure.

%%%%%%%%%%%%%%%%%%%%%%%%%%%%%%%%%%%%%%%%%%%%%%%%%%%%%%%%%%%%


\begin{thebibliography}{99}

\changefontsizes{9pt}


\bibitem{chi_brown}
\textbf{Brown, F.\ Crawford (1931)}:
    ``The eclipse in China'',
    \textit{Popular Astronomy 39}, Dec 1931, p567--573


\bibitem{chi_chalmers}
\textbf{Chalmers, John (1861)}:
    ``Appendix on the Astronomy of the Ancient Chinese'',
    in: \textit{Chinese Classics vol. III: The Shoo King},
    ed.\ by James Legge, Hongkong, 1861, p101--102;
    see \cite{chi_legge}


\bibitem{chi_chambers}
\textbf{Chambers, George F. (1909)}:
    ``The Story of Eclipses'',
    McClure Publ.\ Co., New York, 1909, p65--69


\bibitem{chi_espenak}
\textbf{Espenak, Fred (2006)}:
    ``Predictions for Solar and Lunar Eclipses'',
    Eclipse maps courtesy of Fred Espenak, NASA/Goddard Space Flight Center,\\
\texttt{\small http://www.eclipsewise.com/}


\bibitem{chi_fairbank}
\textbf{Fairbank, J.K. \& Goldman M.\ (2006)}:
    ``China: A new History'', 2nd ed.,
    Harvard University Press, Cambridge/Massachusetts, 2006, p35;
    ISBN 978-0-67401828-0


\bibitem{chi_fotheringham_1921}
\textbf{Fotheringham, John K. (1921)}:
    ``Historical Eclipses'',
    Clarendon Press, Oxford, 1921, p9--16


\bibitem{chi_fotheringham_1933}
\textbf{Fotheringham, John K. (1933)}:
    ``The Story of Hi and Ho'',
    \textit{Journal of the British Astronomical Association 43},
    1933, p248--257


\bibitem{gaubil}
\textbf{Gaubil, Antoine (1732)}:
    ``Trait\'e de l'Astronomie Chinoise'', in:
    \textit{Observations math\'ematiques}, 
    \'Etienne Souciet, Paris, 1729--1732;
    cited by \cite{whiston_1734, chi_rothman_1840}


\bibitem{chi_henriksson_2008}
\textbf{Henriksson, G\"oran (2008)}:
    ``A new Attempt to date the Xia, Shang and Western Zhou Dynasties
        by solar eclipses'',
    \textit{Archeologia Baltica 10}, 2008, p105--109


\bibitem{chi_herrmann}
\textbf{Herrmann, Dieter B. (1998)}:
    ``Die Jahrhundertfinsternis'',
    Paetec Verlag, Berlin, 1998, 1st ed., p22--23 (in German)


\bibitem{chi_keenan_2002}
\textbf{Keenan, Douglas J. (2002)}:
    ``Astro-Historiographic chronologies of early China are unfounded'',
    \textit{East Asian History 23}, 2002, p61--68


\bibitem{finsternisbuch}
\textbf{Khalisi, Emil (2020)}:
%%% MANUELLE SILBENTRENNUNG !
    ``Das Buch der legend\"aren Finster\-nisse'',
    Habilitation submitted to the University of Heidelberg, 2020, ch.\ 10.2
    (in German)


\bibitem{chi_legge}
\textbf{Legge, James (1865)}:
    ``Chinese Classis'', vol. III (The Shoo King),
    Tr\"ubner \& Co., London, 1865


\bibitem{chi_liu_2002}
\textbf{Liu, Ci-yuan (2002)}:
    ``Astronomy in the Xia-Shang-Zhou Chronology Project'',
    \textit{Journal of Astronomical History and Heritage 5}, No. 1/2002, p1--8


\bibitem{chi_liu-etal_2003}
\textbf{Liu C., Liu X. \& Ma L. (2003)}:
    ``Examination of early Chinese records of solar eclipses'',
    \textit{Journal of Astronomical History and Heritage 6}, No. 1/2003, p53--63


\bibitem{chi_nivison-pang}
\textbf{Nivison D.S. \& Pang K.D. (1990)}:
    ``Astronomical Evidence for the Bamboo Annals' Chronicle
        of Early Xia'',
    \textit{Early China 15}, 1990, p87--95
%%% *** siehe Printout ***


\bibitem{chi_oppolzer_1880}
\textbf{Oppolzer, Theodor Ritter von (1880)}:
   ``\"Uber die Sonnenfinsternis des Schu-King'',
   \textit{Monatsbericht der K\"onigl.\ Preu{\ss}.\ Akademie d.W.},
   1880, p166--185 (in German)


\bibitem{chi_pang_1987}
\textbf{Pang, Kevin D. (1987)}:
   ``Extraordinary floods in early Chinese history and their absolute dates'',
   \textit{Journal of Hydrology 96}, 15 Dec 1987, p139--155;\\
   doi: 10.1016/0022-1694(87)90149-1


\bibitem{pang-bangert}
\textbf{Pang K.D. \& Bangert J.A. (1993)}:
    ``The Holy Grail of Chinese Astronomy: \dots '',
    \textit{Bulletin of the American Astronomical Society 25}, 1993, p922


\bibitem{pankenier_1980}
\textbf{Pankenier, David W. (1981)}:
    ``Astronomical Dates in Shang and Western Zhou'',
    \textit{Early China 7}, 1981, p2--37


\bibitem{chi_rothman_1840}
\textbf{Rothman, Richard W. (1840)}:
   ``On an Ancient Solar Eclipse observed in China'',
   \textit{Memoirs of the Royal Astronomical Society 11}, 1840, p47--50


\bibitem{chi_russell_1895}
\textbf{Russell, Samuel M. (1895)}:
    ``Some Astronomical Records in ancient Chinese Books'',
    \textit{The Observatory 18}, 1895 (No.\ 231 + 232 + 234),
       p323--325 + p355--358 + p430--432


\bibitem{chi_schlegel-kuehnert}
\textbf{Schlegel G. \& K\"uhnert F. (1889)}:
    ``Die Schu-King-Finsternis'',
    K\"onigliche Akademie der Wissenschaften,
    Verlag Johannes M\"uller, Amsterdam, 1889 (in German)


\bibitem{chi_shaughnessy_2009}
\textbf{Shaughnessy, Edward L. (2009)}:
    ``Chronologies of Ancient China: A Critique of the
          Xia-Shang-Zhou Chronology Project'',
    in: \textit{Windows on the Chinese World}, ed. by Clara Ho,
    Lexington Books, Lanham/Maryland, 2009, p15--28;
    ISBN 978-0-7391-2769-8


\bibitem{stephenson-morrison_2005}
\textbf{Stephenson, F.R. \& Morrison L.V. (2005)}:
    ``Historical Eclipses'',
    \textit{ASP Conference Series 335}, 2005, p159--180


\bibitem{chi_stockwell_1895}
\textbf{Stockwell, John N. (1895)}:
    ``Eclipses and Chronology'',
    \textit{The Observatory 18}, No. 226/1895, p161--163


\bibitem{chi_wang-siscoe}
\textbf{Wang P.K. \& Siscoe G.L. (1980)}:
%%% Pao-Kuan Wang (Meteorologist), George L. Siscoe (*1937, solar physicist)
    ``Ancient Chinese Observations of Physical Phenomena Attending Solar Eclipses'',
    \textit{Solar Physics 66}, 1980, p187--193;
    doi: 10.1007/BF00150528


\bibitem{weitzel_1945}
\textbf{Weitzel, Robert B. (1945)}:
%%% (Weitzel, Robert?)
    ``Clusters of five planets'',
    \textit{Popular Astronomy 53}, 1945, p159--161


\bibitem{whiston_1734}
\textbf{Whiston, William (1734)}:
    ``Six dissertations'',
    ed.\ by John Whiston, London, 1734, p195--209
    %%% John Whiston (1711--1780) = son
    %%% Josiah Whiston (?) = father


\bibitem{chi_w-en}
\textbf{Wikipedia (2020)}:
    \texttt{https://en.wikipedia.org/ }
    (English + German sites), visited in February 2020


\bibitem{chi_williams_1863}
\textbf{Williams, John (1863)}:
    ``On an Eclipse of the Sun recorded in the Chinese Annals \dots '',
    \textit{Monthly Notices of the Royal Astronomical Society 23}, 06/1863, p238--242;
    doi: 10.1093/mnras/23.8.238


\bibitem{chi_wu-etal_2016}
\textbf{Wu Q., Zhao Z., Liu L. + 13 co-authors (2016)}:
    ``Outburst flood at 1920 BCE supports historicity of \dots '',
    \textit{Science 353}, 5 Aug 2016, p579--582;
    doi: 10.1126/science.aaf0842 
%%% *** siehe Printout ***


\bibitem{wu-etal_2017-response}
\textbf{Wu Q., Zhao Z., Lui L. + 13 co-authors (2017)}:
    ``Response to Comments on \frq Outburst flood at 1920 BCE \dots\flq{}'',
    \textit{Science 355}, 31 Mrc 2017, p1382e;\\
    doi: 10.1126/science.aal1325


\bibitem{zhang-etal_2019}
\textbf{Zhang Y., Huang C.C., Schulmeister J. + 10 co-authors (2019)}:
    ``Formation and evolution of the Holocene massive
        landslide-dammed lakes \dots '',
    \textit{Quaternary Science Reviews 218}, 2019, p267--280;
    doi: 10.1016/j.quascirev.2019.06.011


\end{thebibliography}
\end{document}